\documentclass{aastex631}

\begin{document}

\title{VERITAS discovery of very high energy gamma-ray emission from S3 1227+25 and multiwavelength observations}

\author[0000-0002-2028-9230]{A.~Acharyya}\affiliation{Department of Physics and Astronomy, University of Alabama, Tuscaloosa, AL 35487, USA}
\author[0000-0002-9021-6192]{C.~B.~Adams}\affiliation{Physics Department, Columbia University, New York, NY 10027, USA}
\author{A.~Archer}\affiliation{Department of Physics and Astronomy, DePauw University, Greencastle, IN 46135-0037, USA}
\author[0000-0002-3886-3739]{P.~Bangale}\affiliation{Department of Physics and Astronomy and the Bartol Research Institute, University of Delaware, Newark, DE 19716, USA}
\author[0000-0003-2098-170X]{W.~Benbow}\affiliation{Center for Astrophysics $|$ Harvard \& Smithsonian, Cambridge, MA 02138, USA}
\author[0000-0002-6208-5244]{A.~Brill}\affiliation{N.A.S.A./Goddard Space-Flight Center, Code 661, Greenbelt, MD 20771, USA}
\author{J.~L.~Christiansen}\affiliation{Physics Department, California Polytechnic State University, San Luis Obispo, CA 94307, USA}
\author{A.~J.~Chromey}\affiliation{Center for Astrophysics $|$ Harvard \& Smithsonian, Cambridge, MA 02138, USA}
\author[0000-0002-1853-863X]{M.~Errando}\affiliation{Department of Physics, Washington University, St. Louis, MO 63130, USA}
\author[0000-0002-5068-7344]{A.~Falcone}\affiliation{Department of Astronomy and Astrophysics, 525 Davey Lab, Pennsylvania State University, University Park, PA 16802, USA}
\author[0000-0001-6674-4238]{Q.~Feng}\affiliation{Center for Astrophysics $|$ Harvard \& Smithsonian, Cambridge, MA 02138, USA}
\author{J.~P.~Finley}\affiliation{Department of Physics and Astronomy, Purdue University, West Lafayette, IN 47907, USA}
\author[0000-0002-2944-6060]{G.~M.~Foote}\affiliation{Department of Physics and Astronomy and the Bartol Research Institute, University of Delaware, Newark, DE 19716, USA}
\author[0000-0002-1067-8558]{L.~Fortson}\affiliation{School of Physics and Astronomy, University of Minnesota, Minneapolis, MN 55455, USA}
\author[0000-0003-1614-1273]{A.~Furniss}\affiliation{Department of Physics, California State University - East Bay, Hayward, CA 94542, USA}
\author{G.~Gallagher}\affiliation{Department of Physics and Astronomy, Ball State University, Muncie, IN 47306, USA}
\author[0000-0002-0109-4737]{W.~Hanlon}\affiliation{Center for Astrophysics $|$ Harvard \& Smithsonian, Cambridge, MA 02138, USA}
\author[0000-0002-8513-5603]{D.~Hanna}\affiliation{Physics Department, McGill University, Montreal, QC H3A 2T8, Canada}
\author[0000-0003-3878-1677]{O.~Hervet}\affiliation{Santa Cruz Institute for Particle Physics and Department of Physics, University of California, Santa Cruz, CA 95064, USA}
\author[0000-0001-6951-2299]{C.~E.~Hinrichs}\affiliation{Center for Astrophysics $|$ Harvard \& Smithsonian, Cambridge, MA 02138, USA and Department of Physics and Astronomy, Dartmouth College, 6127 Wilder Laboratory, Hanover, NH 03755 USA}
\author{J.~Hoang}\affiliation{Santa Cruz Institute for Particle Physics and Department of Physics, University of California, Santa Cruz, CA 95064, USA}
\author[0000-0002-6833-0474]{J.~Holder}\affiliation{Department of Physics and Astronomy and the Bartol Research Institute, University of Delaware, Newark, DE 19716, USA}
\author[0000-0002-1089-1754]{W.~Jin}\affiliation{Department of Physics and Astronomy, University of Alabama, Tuscaloosa, AL 35487, USA}
\author[0009-0008-2688-0815]{M.~N.~Johnson}\affiliation{Santa Cruz Institute for Particle Physics and Department of Physics, University of California, Santa Cruz, CA 95064, USA}
\author[0000-0002-3638-0637]{P.~Kaaret}\affiliation{Department of Physics and Astronomy, University of Iowa, Van Allen Hall, Iowa City, IA 52242, USA}
\author{M.~Kertzman}\affiliation{Department of Physics and Astronomy, DePauw University, Greencastle, IN 46135-0037, USA}
\author[0000-0003-4785-0101]{D.~Kieda}\affiliation{Department of Physics and Astronomy, University of Utah, Salt Lake City, UT 84112, USA}
\author[0000-0002-4260-9186]{T.~K.~Kleiner}\affiliation{DESY, Platanenallee 6, 15738 Zeuthen, Germany}
\author[0000-0002-4289-7106]{N.~Korzoun}\affiliation{Department of Physics and Astronomy and the Bartol Research Institute, University of Delaware, Newark, DE 19716, USA}
\author{F.~Krennrich}\affiliation{Department of Physics and Astronomy, Iowa State University, Ames, IA 50011, USA}
\author[0000-0003-4641-4201]{M.~J.~Lang}\affiliation{School of Natural Sciences, University of Galway, University Road, Galway, H91 TK33, Ireland}
\author[0000-0003-3802-1619]{M.~Lundy}\affiliation{Physics Department, McGill University, Montreal, QC H3A 2T8, Canada}
\author[0000-0001-9868-4700]{G.~Maier}\affiliation{DESY, Platanenallee 6, 15738 Zeuthen, Germany}
\author{C.~E~McGrath}\affiliation{School of Physics, University College Dublin, Belfield, Dublin 4, Ireland}
\author[0000-0001-7106-8502]{M.~J.~Millard}\affiliation{Department of Physics and Astronomy, University of Iowa, Van Allen Hall, Iowa City, IA 52242, USA}
\author{J.~Millis}\affiliation{Department of Physics and Astronomy, Ball State University, Muncie, IN 47306, USA and Department of Physics, Anderson University, 1100 East 5th Street, Anderson, IN 46012}
\author[0000-0001-5937-446X]{C.~L.~Mooney}\affiliation{Department of Physics and Astronomy and the Bartol Research Institute, University of Delaware, Newark, DE 19716, USA}
\author[0000-0002-1499-2667]{P.~Moriarty}\affiliation{School of Natural Sciences, University of Galway, University Road, Galway, H91 TK33, Ireland}
\author[0000-0002-3223-0754]{R.~Mukherjee}\affiliation{Department of Physics and Astronomy, Barnard College, Columbia University, NY 10027, USA}
\author[0000-0002-9296-2981]{S.~O'Brien}\affiliation{Physics Department, McGill University, Montreal, QC H3A 2T8, Canada}
\author[0000-0002-4837-5253]{R.~A.~Ong}\affiliation{Department of Physics and Astronomy, University of California, Los Angeles, CA 90095, USA}
\author[0000-0001-7861-1707]{M.~Pohl}\affiliation{Institute of Physics and Astronomy, University of Potsdam, 14476 Potsdam-Golm, Germany and DESY, Platanenallee 6, 15738 Zeuthen, Germany}
\author[0000-0002-0529-1973]{E.~Pueschel}\affiliation{DESY, Platanenallee 6, 15738 Zeuthen, Germany}
\author[0000-0002-4855-2694]{J.~Quinn}\affiliation{School of Physics, University College Dublin, Belfield, Dublin 4, Ireland}
\author[0000-0002-5351-3323]{K.~Ragan}\affiliation{Physics Department, McGill University, Montreal, QC H3A 2T8, Canada}
\author{P.~T.~Reynolds}\affiliation{Department of Physical Sciences, Munster Technological University, Bishopstown, Cork, T12 P928, Ireland}
\author[0000-0002-7523-7366]{D.~Ribeiro}\affiliation{School of Physics and Astronomy, University of Minnesota, Minneapolis, MN 55455, USA}
\author{E.~Roache}\affiliation{Center for Astrophysics $|$ Harvard \& Smithsonian, Cambridge, MA 02138, USA}
\author[0000-0003-1387-8915]{I.~Sadeh}\affiliation{DESY, Platanenallee 6, 15738 Zeuthen, Germany}
\author{A.~C.~Sadun}\affiliation{Department of Physics, University of Colorado Denver, Campus Box 157, P.O.Box 173364, Denver CO 80217, USA}
\author[0000-0002-3171-5039]{L.~Saha}\affiliation{Center for Astrophysics $|$ Harvard \& Smithsonian, Cambridge, MA 02138, USA}
\author{M.~Santander}\affiliation{Department of Physics and Astronomy, University of Alabama, Tuscaloosa, AL 35487, USA}
\author{G.~H.~Sembroski}\affiliation{Department of Physics and Astronomy, Purdue University, West Lafayette, IN 47907, USA}
\author[0000-0002-9856-989X]{R.~Shang}\affiliation{Department of Physics and Astronomy, Barnard College, Columbia University, NY 10027, USA}
\author[0000-0003-3407-9936]{M.~Splettstoesser}\affiliation{Santa Cruz Institute for Particle Physics and Department of Physics, University of California, Santa Cruz, CA 95064, USA}
\author{A.~Kaushik~Talluri}\affiliation{School of Physics and Astronomy, University of Minnesota, Minneapolis, MN 55455, USA}
\author{J.~V.~Tucci}\affiliation{Department of Physics, Indiana University-Purdue University Indianapolis, Indianapolis, IN 46202, USA}
\author{V.~V.~Vassiliev}\affiliation{Department of Physics and Astronomy, University of California, Los Angeles, CA 90095, USA}
\author[0000-0003-2740-9714]{D.~A.~Williams}\affiliation{Santa Cruz Institute for Particle Physics and Department of Physics, University of California, Santa Cruz, CA 95064, USA}
\author[0000-0002-2730-2733]{S.~L.~Wong}\affiliation{Physics Department, McGill University, Montreal, QC H3A 2T8, Canada}
\collaboration{64}{(The VERITAS Collaboration)}

\author[0000-0002-2024-8199]{Talvikki Hovatta}\affiliation{Finnish Centre for Astronomy with ESO (FINCA), University of Turku, FI-20014, Turku, Finland}  \affiliation{Aalto University Mets\"ahovi Radio Observatory, Mets\"ahovintie 114,02540 Kylm\"al\"a, Finland}
\author[0000-0001-6158-1708]{Svetlana G. Jorstad}\affiliation{Institute for Astrophysical Research, Boston University, 725 Commonwealth Avenue, Boston, MA 02215, USA}
\author[0000-0001-6314-9177]{Sebastian Kiehlmann} \affiliation{Institute of Astrophysics, Foundation for Research and Technology-Hellas, GR-71110 Heraklion, Greece} \affiliation{Department of Physics, Univ. of Crete, GR-70013 Heraklion, Greece}
\author[0000-0002-0393-0647]{Anne L\"ahteenm\"aki}\affiliation{Aalto University Mets\"ahovi Radio Observatory, Mets\"ahovintie 114,02540 Kylm\"al\"a, Finland} \affiliation{Aalto University Department of Electronics and Nanoengineering, P.O. BOX 15500, FI-00076 AALTO, Finland} 
\author[0000-0001-9200-4006]{Ioannis Liodakis} \affiliation{Finnish Center for Astronomy with ESO, University of Turku, FI-20014, Finland}
\author[0000-0001-7396-3332]{Alan P. Marscher}\affiliation{Institute for Astrophysical Research, Boston University, 725 Commonwealth Avenue, Boston, MA 02215, USA}
\author[0000-0002-5491-5244]{Walter Max-Moerbeck} \affiliation{Departamento de Astronom\'ia, Universidad de Chile, Camino El Observatorio 1515, Las Condes, Santiago, Chile}
\author[0000-0001-9152-961X]{Anthony C. S. Readhead} \affiliation{Owens Valley Radio Observatory, California Institute of Technology, Pasadena, CA 91125, USA}
\author[0000-0001-5704-271X]{Rodrigo Reeves} \affiliation{Departamento de Astronom\'a, Universidad de Concepti\'on, Concepci\'on, Chile}
\author[0000-0002-5083-3663]{Paul S. Smith}\affiliation{Steward Observatory, University of Arizona, 933 North Cherry Avenue, Tucson, AZ 85721-0065, USA}
\author[0000-0003-1249-6026]{Merja Tornikoski}\affiliation{Aalto University Mets\"ahovi Radio Observatory, Mets\"ahovintie 114,02540 Kylm\"al\"a, Finland}
\nocollaboration{11}

\correspondingauthor{Atreya Acharyya}
\email{aacharyya1@ua.edu}

\correspondingauthor{Marcos Santander}
\email{jmsantander@ua.edu}



\begin{abstract}

We report the detection of very high energy gamma-ray emission from the blazar S3 1227+25 (VER J1230+253) with the Very Energetic Radiation Imaging Telescope Array System (VERITAS). 
VERITAS observations of the source were triggered by the detection of a hard-spectrum GeV flare on May 15, 2015 with the \emph{Fermi}-Large Area Telescope (LAT).
A combined five-hour VERITAS exposure on May 16th and May 18th resulted in a strong $13\sigma$ detection  with a differential photon spectral index, $\Gamma = $ 3.8 $\pm$ 0.4, and a flux level at 9\% of the Crab Nebula above 120 GeV. This also triggered target of opportunity observations with \emph{Swift}, optical photometry, polarimetry and radio measurements, also presented in this work, in addition to the VERITAS and \emph{Fermi}-LAT data. A temporal analysis of the gamma-ray flux during this period finds evidence of a shortest variability timescale of $\tau_{\text{obs}}=6.2 \pm 0.9$ hours, indicating emission from compact regions within the jet, and the combined gamma-ray spectrum shows no strong evidence of a spectral cut-off. An investigation into correlations between the multiwavelength observations found evidence of optical and gamma-ray correlations, suggesting a single-zone model of emission. Finally, the multiwavelength spectral energy distribution is well described by a simple one-zone leptonic synchrotron self-Compton radiation model.

\end{abstract}



\section{Introduction}
\label{sec:intro}

The extragalactic gamma-ray sky is dominated by blazars~\citep{gammaagn}, a subclass of radio-loud active galactic nuclei (AGN) powered by a central supermassive black hole (SMBH), that display relativistic jets pointed close to the Earth. Blazars are characterized by non-thermal continuum emission spanning the entire electromagnetic spectrum and by very fast variability which has been observed down to timescales of minutes in the very high energy (VHE, $E_{\gamma}\geq$ 100 GeV) regime~\citep{1996_Gaidos,Mrk_501_minute_variability, PKS_2155_minute_variability, 2020_Mrk421}. 

The spectral energy distribution (SED) of blazars typically displays two broad emission features. The first one, in the radio to X-ray waveband, is commonly attributed to synchrotron emission from relativistic electrons and positrons in the jet. The origin of the second feature, in the X-ray to gamma-ray band, is less clear and two main scenarios have been postulated to explain it. In leptonic models, the emission is attributed to inverse Compton (IC) scattering between the energetic leptons in the jet and a field of low energy photons (for example \citealt{1974_Jones, 1994_Sikora}), either the same synchrotron photons (synchrotron self-Compton [SSC] model), or from an external source (external inverse Compton [EIC] model) such as the accretion disk, the broad-line region (BLR), or the molecular torus (MT). In hadronic scenarios, the high energy photons are produced in  cosmic-ray interactions through the decay of neutral pions (for example \citealt{1992_Mannheim}) or proton synchrotron emission (for example \citealt{2002_Aharonian}). 

The strong radio source S3 1227+25~\citep{1972AJ.....77..797P}, also known as ON 246 \citep{1968AJ.....73..381D}, was first identified as a candidate BL Lacertae (BL Lac) object in a correlation study between the ROSAT All-Sky Survey and the Hamburg Quasar Survey~\citep{1994A&A...286..381B}, which selected sources with strong X-ray to optical flux ratios. 
BL Lacs, a subclass of blazars that display weak emission lines, are typically classified as low (LBL), intermediate (IBL) or high (HBL) synchrotron-peaked objects based on the frequency of the synchrotron peak in their SED \citep{1995MNRAS.277.1477P, 2006A&A...445..441N, 2010ApJ...716...30A}. Several values are reported in the literature for the synchrotron peak frequency ($\nu_{\mathrm{peak}}$) of S3 1227+25: \cite{2009RAA.....9..168W} find a value of  $10^{14.11}$ Hz by fitting archival measurements of the synchrotron portion of the SED, while \cite{2016ApJS..226...20F} report $10^{14.42\pm0.13}$ Hz, and the Third \emph{Fermi}-LAT AGN catalog lists $10^{14.91}$ Hz~\citep{2015ApJ...810...14A}, placing this source near the boundary between IBL and HBL objects, with only 9 IBL objects detected so far in the VHE band.\footnote{\url{http://tevcat2.uchicago.edu/} (accessed on 02/27/2023)}

Radio observations of S3 1227+25 with VLBI~\citep{2008MNRAS.384..230K} have resolved a pc-scale core and jet with apparent bending. The polarization vectors are roughly aligned in the direction of the jet in its inner region, indicating a transverse magnetic field, and a sheath of longitudinal magnetic field lines along one of the outer edges of the jet. 
\cite{2021_Lister} have reported a maximum proper motion in the jet of 0.45 $\pm$ 0.04~mas/yr which translates into an apparent speed of $\sim$4c assuming a redshift $z = 0.135$ \citep{1996A&A...309..419N}, although the redshift of the source is uncertain. 

%
%

%
%
\cite{1996A&A...309..419N} report a redshift $z = 0.135$, and this value has been cited in most catalogs that include S3 1227+25 (for example \citealt{2015ApJ...810...14A, goldoni_paolo_2021_4721386}).  However, \cite{1996A&A...309..419N} do not show a spectrum for S3 1227+25, and for objects where they do not show a new spectrum, they use the spectral results from \cite{1994A&A...286..381B}. \cite{1994A&A...286..381B} show a featureless spectrum for S3 1227+25, but have another object with absorption lines from the host galaxy at redshift, $z = 0.135$, RX J11365+6737.  Table~3 of \cite{1996A&A...309..419N} includes both S3 1227+25, with redshift $z = 0.135$, and RX J11365+6737, with no redshift.  We conclude that this is a typographical error in the table and that the redshift of RX J11365+6737 was entered in the wrong row.  Neither \cite{1996A&A...309..419N} nor \cite{1994A&A...286..381B} show a spectrum from which the redshift of S3 1227+25 can be determined. We obtained new spectroscopy of S3 1227+25 on February 12, 2016 (MJD 57430) using the Kast double spectrograph on the 3 m Shane telescope at the Lick Observatory for 30 minutes with red and blue sides covering the 3500-7000~\AA~wavelength range. The spectrum was dominated by the continuum emission, and no clear indications of the typical features of a host galaxy (such as G band, Ca H\&K or continuum break, Mg Ib, or Na ID) were visible.

\begin{figure}
\centering
\includegraphics[width= 0.6 \linewidth]{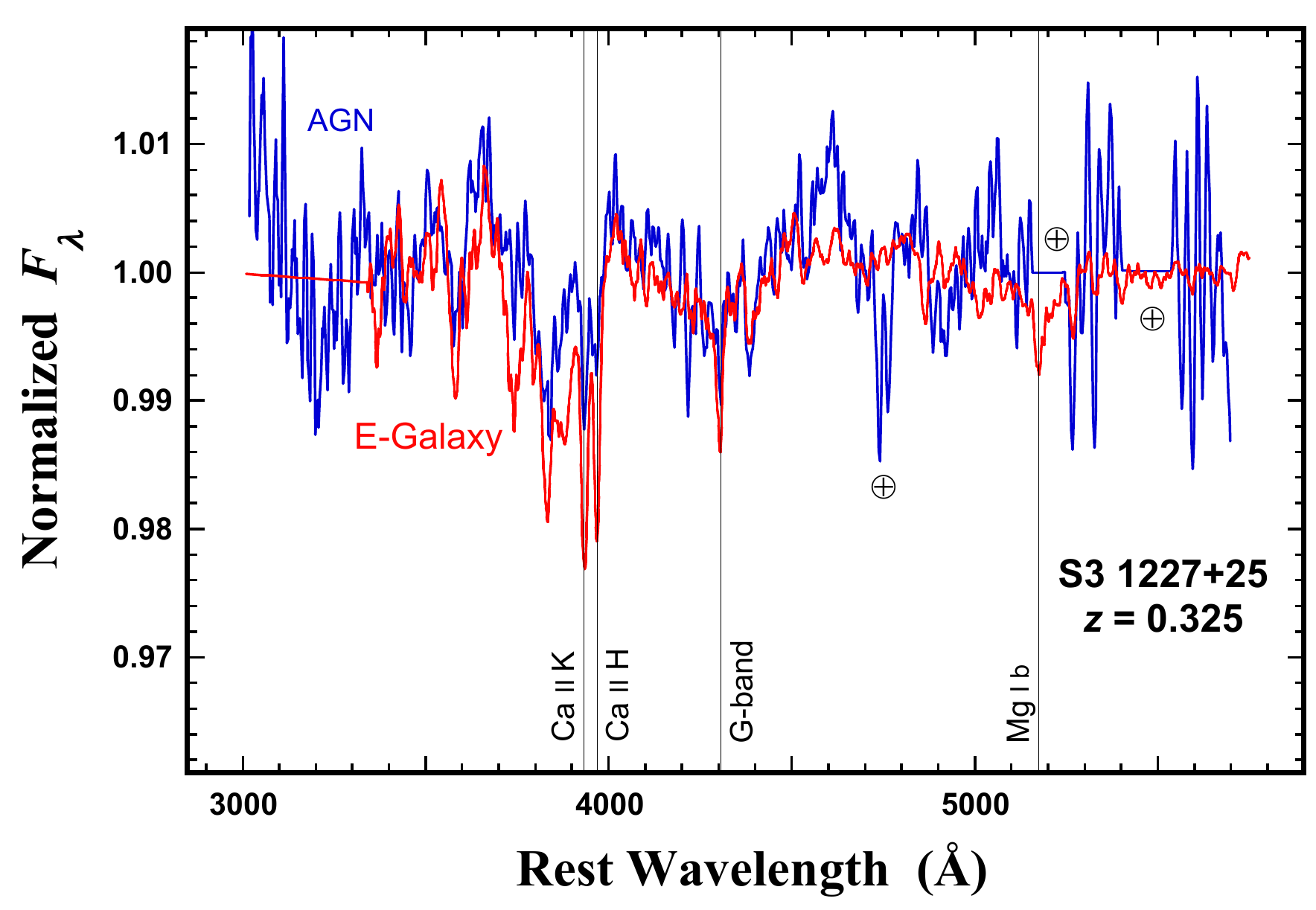}
\caption{The normalized median spectrum of S3 1227+25 compared to the normalized spectrum of a typical late-type galaxy. Identifying observed features with Ca II H\&K and the G-band yielded a redshift of $z=0.325$. Major telluric absorption features in the spectrum of the AGN are marked with the terrestrial symbol. The redder telluric features have been edited out of the spectrum of S3 1227+25.}
\label{fig:opt_spectrum}
\end{figure}

Other recent spectroscopy (for example~\citealt{2015A&A...575A.124M, 2017ApJ...837..144P}) has also failed to determine the redshift. Redshift limits, $0.32 \leq z \leq 1.78$, have been obtained by \cite{2013ApJ...764..135S} as part of a large spectroscopic survey of BL Lac sources, where the lower limit depends on an assumed absolute magnitude for the host galaxy and the upper limit is determined statistically from the absence of Ly$\alpha$ absorption features. The normalized median spectrum of S3 1227+25, shown in Fig.~\ref{fig:opt_spectrum}, is from the Steward Observatory blazar monitoring program.\footnote{Credit: Paul S. Smith, full paper in preperation.} Also shown for comparison is the normalized spectrum of a typical late-type galaxy. There is good correspondence between features between 3400--4500 \AA, covering the region around the H\&K break. Identifying observed features with Ca II H\&K and the G-band yielded a redshift of $z=0.325$. The expected elliptical galaxy Mg Ib feature is hidden in S3 1227+25 by the O2 B-band absorption feature. Based on the above discussion, a redshift $z=0.325$, was assumed for the scientific results presented in this work.


A gamma-ray source positionally consistent with S3~1227+25 was first identified in \emph{Fermi}-LAT data and is listed in the \emph{Fermi}-LAT one-year general point source catalog and subsequent updates, most recently in the 12-year catalog (4FGL-DR3, \citealt{4fgl_dr3}). The source was also included in the first \emph{Fermi}-LAT catalog of sources detected above 10 GeV (1FHL, \citealt{2013ApJS..209...34A}), although it was not identified as a potential VHE emitter given its low flux (about 1\% of the Crab Nebula flux above 10 GeV) and a relatively soft, albeit uncertain, spectral index, $\Gamma_{\mathrm{1FHL}} = 3.29 \pm 0.95$. While the blazar was not listed in the second catalog of \emph{Fermi}-LAT sources detected above 50~GeV (2FHL, \citealt{2016ApJS..222....5A}), it was included in the third \emph{Fermi}-LAT catalog  of sources significantly detected in the 10 GeV -- 2 TeV energy range (3FHL, \citealt{3fhl}), with a spectral index, $\Gamma_{\mathrm{3FHL}} = 2.69 \pm 0.22$. Since its first detection, several hard-spectrum GeV flares from S3 1227+25 have been observed with the \emph{Fermi}-LAT~\citep{2015ATel.7596....1B, 2015ATel.6982....1C}. 

This work reports the follow-up to the discovery of VHE emission from S3 1227+25 with VERITAS~\citep{2015ATel.7516....1M}, based on observations initiated on May 16, 2015 UTC (MJD 57158), following the detection of a strong, hard-spectrum flare of the source with the \emph{Fermi}-LAT on May~15. The VERITAS detection triggered target of opportunity (ToO) observations with \emph{Swift}, optical photometry, polarimetry, and radio measurements, which are also presented. In particular, we aim to constrain the characteristics and location of the emission region using the large photon statistics during the VHE detection period. In Section \ref{obs} we discuss the observations and data analysis routines, from high to low frequencies (VERITAS, \emph{Fermi}-LAT, X-ray observations, optical telescope data and radio observations). In Section \ref{discussion} we describe the methods used to constrain the size and location of the emission region during the VHE detection period. We summarise our conclusions in Section \ref{sec:conclusions}.

\section{Observations and Data Analysis}
\label{obs}
\subsection{VERITAS} \label{sec:vts}

The Very Energetic Radiation Imaging Telescope Array System (VERITAS) is an array of four 12 m imaging atmospheric Cherenkov telescopes (IACTs) located at the Fred Lawrence Whipple Observatory (FLWO) in southern Arizona, USA ($30^{\circ}$ 40' N, $110^{\circ}$ 57' W, 1.3 km above sea level, \citealt{vts_paper}).  Each telescope is equipped with a camera containing 499 photomultiplier tubes, covering a field of view of 3.5$^{\circ}$ and is capable of detecting gamma-rays having energies from 85 GeV to above 30 TeV, with an energy resolution of $ \Delta E/E~\sim$~15\% (at 1~TeV) and an angular resolution of $\sim$~0.1$^{\circ}$ (68\% containment at 1~TeV). In its current configuration, VERITAS typically detects a source with a flux of 1\% of the steady state gamma-ray flux of the Crab Nebula, at a statistical significance of $>5\sigma$, in 25 hours of observation. It should be noted that the required time to reach a $5\sigma$ detection is longer for sources with a spectrum softer than the Crab Nebula (photon index, $\Gamma > 2.49$) or those observed at large zenith angles ($\theta_{z} > 40^{\circ}$) \citep{Park15}.

VERITAS observations of S3 1227+25 were started on May~16, 2015 UTC (MJD 57158), following the report of a hard-spectrum GeV flare with the \emph{Fermi}-LAT. The source was observed for 6.9 hours between May~16 and May~23 (MJD 57158 - MJD 57165), at an average zenith angle of $14^{\circ}$, with bad weather conditions at the VERITAS site occasionally preventing observations during this period. 
The observations were performed using a standard ``wobble'' observation mode \citep{FOMIN1994151} with a $0.5^{\circ}$ offset in each of the four cardinal directions to allow for a simultaneous determination of the background event rate. Quality cuts were applied to the data set to remove observation periods affected by bad weather. The VERITAS data were analyzed using the Eventdisplay analysis package \citep{Maier17}, and independently confirmed with  the VEGAS analysis package \citep{VEGAS}, yielding consistent results.

The VERITAS analysis parametrizes the principal moments of the elliptical shower images and applies a set of cuts to these parameters to reject cosmic-ray background events. The cuts are determined using a boosted decision tree algorithm~\citep{2017APh....89....1K}, previously trained on gamma-ray shower simulations and optimized for soft-spectrum sources ($\Gamma \sim 4$). Only events with at least two telescope images are selected during this process. Gamma-ray candidate events are considered to fall in the source (\textsc{ON}) region if the squared angular distance, $\theta^{2}$, between the reconstructed event origin and S3 1227+25 is less than 0.008 deg$^{2}$. The background is estimated using the reflected region model \citep{2007A&A...466.1219B} where circular regions with the same size as the \textsc{ON} region are placed at the same radial distance from the center of the camera. 
The source was detected on two separate nights, May~16 (MJD 57158) and May~18 (MJD 57160), during which a total of 4.9 hours of observations were obtained. Bad weather meant S3 1227+25 could not be observed with VERITAS on May~17 (MJD 57159).

\begin{figure}
\centering
\includegraphics[width=0.4\linewidth]{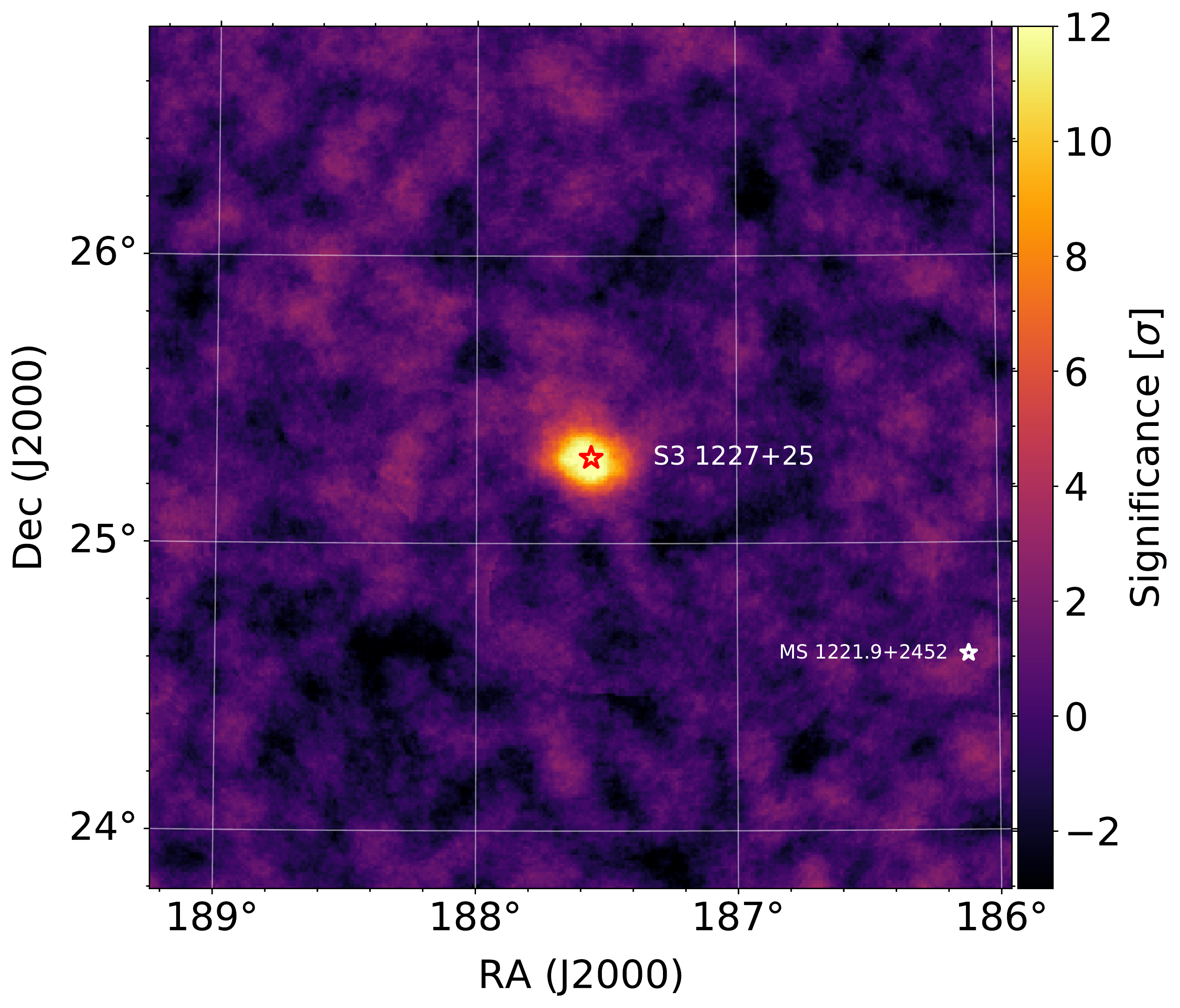}
\includegraphics[width=0.5 \linewidth]{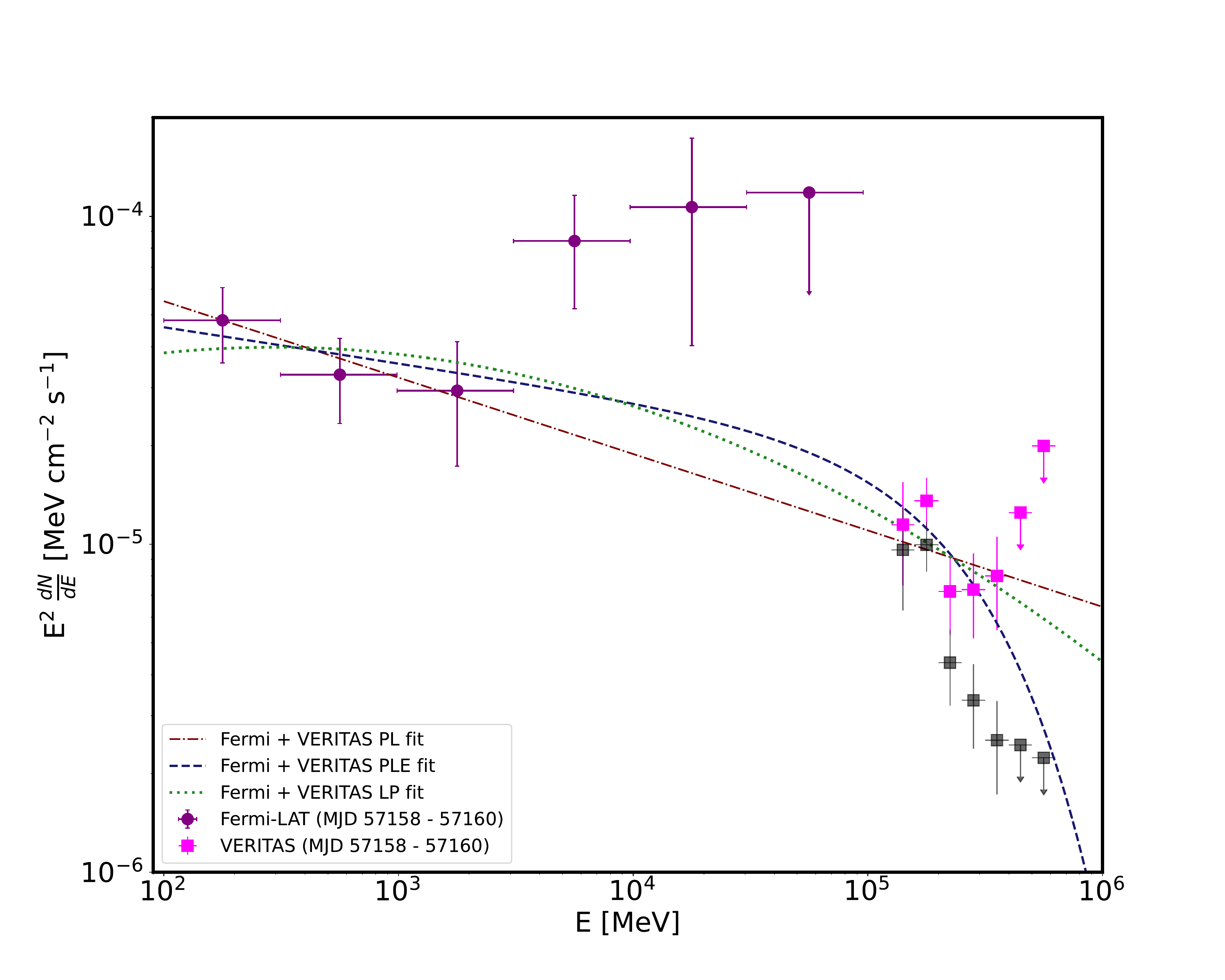}
\caption{Left: VERITAS significance skymap based on observations taken on May 16, 2015 and May 18, 2015 UTC (MJD 57158 and 57160) at the location of the new VHE source positionally coincident with S3 1227+25. The known VHE blazar MS 1221.9+2452, shown with a white marker, was not detected during this period. Right: The purple data points show the \textit{Fermi}-LAT spectrum obtained for S3 1227+25 in the energy range 100~MeV--100~GeV between May 16 and May 18, 2015 UTC (MJD 57158 -- 57160). The data are binned into two energy bins per decade, with individual bins having a TS $<$ 4 (roughly corresponding to a significance of $\sim$ 2$\sigma$) considered as upper limits. The magenta data points show the VHE spectrum for S3 1227+25 based on VERITAS observations taken on May 16 and May 18, 2015 UTC (MJD 57158 -- 57160). A minimum significance of two standard deviations and three on-source photons are required for each flux point and upper limits at 95\% confidence level are quoted otherwise.  De-absorbed flux points, using the \cite{RN10} EBL model, are shown in magenta and the corresponding observed fluxes are shown in black. A power law (PL), power law with exponential cut-off (PLE) and log parabola (LP) function is fit to the de-absorbed combined gamma-ray spectrum and is shown as dashed red, blue and green curves respectively.}
\label{fig:combined_spectrum}
\end{figure}

An excess of 193 gamma-ray candidate events were recorded in the source region with $N_{\text{ON}}$ = 333 \textsc{ON} events, $N_{\text{OFF}}$ = 1164 \textsc{OFF} events, a background normalization factor, $\alpha = 0.12$, and $N_{\text{Excess}} = N_{\text{ON}} - \alpha N_{\text{OFF}} = 193$, corresponding to a statistical significance of 12.7$\sigma$, calculated following the method of \cite{1983ApJ...272..317L}. The background normalization factor, $\alpha$, is defined as the ratio of the on-source exposure to the background exposure and accounts for differences, for example in solid angle, exposure time, zenith angle and the detector acceptance for gamma-ray like cosmic-ray events, between the the \textsc{ON} region and the background region \citep{2007A&A...466.1219B}.
A significance skymap centered on the position of S3 1227+25 is shown in Fig.~\ref{fig:combined_spectrum}. The excess event map was fit with a two-dimensional Gaussian function to determine the source centroid, found at a distance of $1.3' \pm 0.4'_{\mathrm{stat}} \pm 0.8'_{\mathrm{sys}}$ from the VLBA location of S3 1227+25, $\alpha_{2000} = 12^{\mathrm{h}} \; 30 \arcmin \; 14.09  \arcsec $ and $\delta_{2000} = +25^{\circ} \; 18 \arcmin \; 7.14  \arcsec$ \citep{2002ApJS..141...13B}, within statistical and systematic uncertainties. S3 1227+25 is therefore assigned the VERITAS catalog name VER~J1230+253.


The differential energy spectrum, shown in Fig.~\ref{fig:combined_spectrum}, was derived using observations on May~16 and May~18 (MJD 57158 - MJD 57160) and fit with a power law function of the form $dN/dE = I_0 (E/E_{0})^{-\Gamma}$ where $E_{0}$ = 0.2 TeV. A good fit ($\chi^2 / \mathrm{ndf}$ = 2.6/3) is found for a flux normalization $I_0 = (1.61 \pm 0.19_{\mathrm{stat}}) \times 10^{-10}$ cm$^{-2}$ s$^{-1}$ TeV$^{-1}$ and a spectral index $\Gamma = 3.79 \pm 0.40_{\mathrm{stat}}$. The average integral flux above 0.12 TeV  during the two nights was $(4.51 \pm 0.44) \times 10^{-11}$ cm$^{-2}$ s$^{-1}$, or 8.9\% of the Crab Nebula flux above the same energy threshold. 
Furthermore, while observations on May 19, 20 and 23 (MJD 57161, 57162, and 57165 respectively) did not yield a detection, 95\% confidence level upper limits on the gamma-ray flux of the source were derived during those periods. A lightcurve of the VERITAS flux above a threshold of 120~GeV and binned in daily intervals is shown in Fig.~\ref{fig:mwl_lc}. Finally, we found no evidence of intra-night variability in the VERITAS data as a lightcurve for observations taken on May 16 and May 18, binned in 4 min intervals, was found to be well fit by a constant function, with a $\chi^{2}/\text{n.d.f}=0.76$, indicating that a constant flux is consistent with the data given the statistical uncertainties.

\subsection{{\it Fermi}-LAT} 
\label{sec:fermi}
The \emph{Fermi}-Large Area Telescope (LAT; \citealt{Fermi_LAT}) is a pair conversion telescope capable of detecting gamma-ray photons in the energy range between 20~MeV to above 500 GeV. Primarily operating in survey mode, the \textit{Fermi}-LAT scans the entire sky every three hours. In this work, we analyzed \textit{Fermi}-LAT photons detected between MJD~57128 and MJD~57189, which corresponds to midnight on April~16, 2015 until midnight on June~16, 2015, a month either side of the VERITAS detection~\citep{2015ATel.7516....1M}. Throughout the analysis, we use the \textit{Fermi} Science Tools version $11-05-03$,\footnote{\url{http://fermi.gsfc.nasa.gov/ssc/data/analysis/software} (accessed on 02/27/2023)} \textit{FERMIPY} version 1.0.1~\footnote{\url{http://fermipy.readthedocs.io} (accessed on 02/27/2023)} \citep{wood2017fermipy} in conjunction with the latest \textit{PASS} 8 IRFs~\citep{atwood2013pass}.

Photons with energies between 100 MeV and 100 GeV that were detected within a region of interest (RoI) of radius 15$^{\circ}$ centered on the location of S3 1227+25 were selected for the analysis.
Furthermore, we selected only photon events from within a maximum zenith angle of 90$^{\circ}$ in order to reduce contamination from background photons from the Earth's limb, produced from the interaction of cosmic-rays with the upper atmosphere.

The contributions from the isotropic and Galactic diffuse backgrounds were modeled using the most recent templates for isotropic and Galactic diffuse emission, iso\_P8R3\_SOURCE\_V2\_v1.txt and gll\_iem\_v07.fits respectively.
Sources in the 4FGL-DR3 catalog within a radius of $20^{\circ}$ from the source position of S3 1227+25 were included in the model with their spectral parameters fixed to their catalog values.
This takes into account the gamma-ray emission from sources lying outside the RoI which might yet contribute photons to the data, especially at low energies, due to the size of the point spread function of the \textit{Fermi}-LAT.

The normalization factor for both the isotropic and Galactic diffuse emission templates were left free along with the spectral normalization of all modeled sources within the RoI. Moreover, the spectral shape parameters of all modeled sources within 3$^{\circ}$ of S3 1227+25 were left free to vary while those of the remaining sources were fixed to the values reported in the 4FGL-DR3 catalog \citep{4fgl_dr3}. 
A binned likelihood analysis was then performed in order to obtain the spectral parameters best describing the model during the period of observation, using a spatial binning of 0.1$^{\circ}$ pixel$^{-1}$ and two energy bins per decade.

The source was detected at a high statistical significance of more than 49$\sigma$ (TS = 2476.7) during the observation period analyzed. 
The spectrum was found to be best modeled by a log parabola (LP):
\begin{equation}
    \hspace*{3cm}
    \centering
   \frac{dN}{dE}=N_{0} \left(\frac{E}{E_0}\right)^{-\alpha - \beta \text{ln} \left(\frac{E}{E_0}\right)}
	\label{LP_spectrum}
\end{equation}
where $N_{0}$ is the normalization, $E_{0}$ is the pivot energy, $\alpha$ the spectral index and $\beta$ the curvature.
The best-fit spectral parameters obtained for S3 1227+25 during the period investigated are $N_{0} = (5.04 \pm 0.27) \times 10^{-11}$ $\text{cm}^{-2}\text{s}^{-1} \text{MeV}^{-1}$, $\alpha = 1.85 \pm 0.05$ and $\beta = 0.05 \pm 0.02$. For comparison, the 4FGL-DR3 catalog spectral values for this source are $N_{0} = (7.21 \pm 0.16) \times 10^{-12}$ $\text{cm}^{-2}\text{s}^{-1} \text{MeV}^{-1}$, $\alpha = 2.02 \pm 0.02$ and $\beta = 0.06 \pm 0.01$.

\begin{figure*}[t!]
\centering
\includegraphics[width=  0.95 \linewidth]{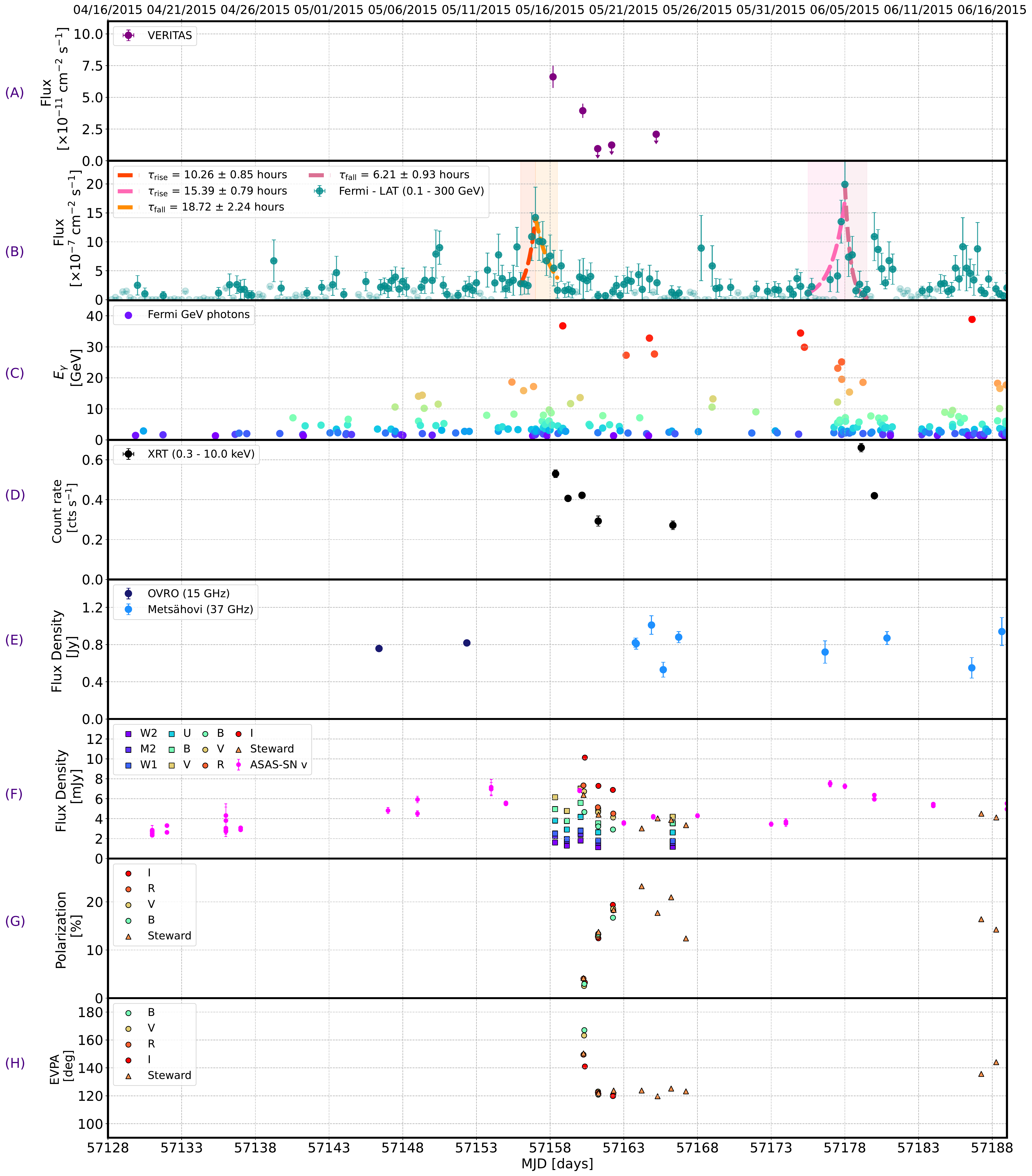}
\caption{Multiwavelength lightcurves of S3 1227+25 between MJD~57128 and MJD~57189, which corresponds to midnight on April~16, 2015 until midnight on June~16, 2015, a month either side of the VERITAS detection. \textbf{(A)} Daily-binned VHE lightcurve for VERITAS observations above an energy threshold of 120 GeV. Upper limits at 95\% confidence level are quoted for observations with a significance of lower than $2\sigma$. \textbf{(B)} \emph{Fermi}-LAT lightcurve in 6 hour bins in the 0.1--300 GeV band. The shaded regions indicate the two periods of flaring activity used to compute the variability timescales, obtained from the fits (dashed curves) and shown in legend. Upper limits are shown for individual bins having a significance of lower than $2\sigma$. \textbf{(C)} \emph{Fermi}-LAT GeV photon energies positionally coincident (at 95\% containment) with the source position. \textbf{(D)} \emph{Swift}-XRT count rate in the 0.3--10 keV band. \textbf{(E)} Radio flux observations: OVRO (dark blue) and Mets\"{a}hovi (light blue) lightcurves at 15 GHz and 37 GHz respectively. \textbf{(F)} Optical/UV flux observations: \emph{Swift}-UVOT (square markers), Lowell Observatory (circles), Steward photometry in the 4000-7000~\AA~ band (triangles) and optical data, not corrected for extinction, from the ASAS-SN Sky Patrol (magenta points). \textbf{(G)} Electric vector polarization angle (EVPA): Lowell Observatory (circles), Steward photometry in the 4000-7000 \AA~ band (triangles).  \textbf{(H)}: Polarization: Lowell Observatory (circles), and Steward photometry in the 4000-7000 \AA~ band (triangles).}
\label{fig:mwl_lc}
\end{figure*}

The differential energy spectrum, shown in Fig.~\ref{fig:combined_spectrum}, was derived using \textit{Fermi}-LAT observations between May~16 and May~18, 2015 (MJD 57158 -- 57160). This spectrum was not fitted with a log parabolic model, shown in Equation~\ref{LP_spectrum}, due to a physically unrealistic negative value obtained for the spectral curvature, $\beta$, and instead we chose to model it using a simple power law (PL) model of the form $\frac{dN}{dE}=N_{0} \left(\frac{E}{E_0}\right)^{-\gamma}$. The best-fit spectral parameters are the normalization $N_{0} = (6.52 \pm 1.12) \times 10^{-11}$ $\text{cm}^{-2}\text{s}^{-1} \text{MeV}^{-1}$ and spectral index $\gamma = 1.79 \pm 0.14$. A lightcurve of the integral flux above 100 MeV, shown in Fig.~\ref{fig:mwl_lc}, is calculated with 6 hour bins and keeping the spectral index fixed. 

Furthermore, we investigated the GeV photon emission from S3 1227+25 for the time interval corresponding to the VERITAS detection. For the purpose of this study, GeV photons are defined as photons having an energy $E_{\gamma} \geq$~1~GeV in the rest frame of the host galaxy. The \textit{Fermi} science  tool \textit{gtsrcprob} estimates the probability of each photon being associated with each source in the RoI and was used to check that the GeV photons detected are associated with S3~1227+25. Before this step, it was necessary to first account for the diffuse components using the \textit{Fermi}  science  tool \textit{gtdiffrsp} and the response was added to the input data. 

Fig. \ref{fig:mwl_lc} (C) shows the GeV photons emitted from within a radius of 0.1$^{\circ}$ around S3 1227+25 and having a $\geq$~95~$\%$ probability of originating from it. While the most energetic photon from S3 1227+25 during the period investigated, having energy 33.4~$\pm$~2.1~GeV, was emitted on MJD 57186.63 and is outside of the time window of the VERITAS observations, the second most energetic photon, having energy 31.6~$\pm$~1.9~GeV, was emitted on MJD 57158.86 and coincides with the VHE discovery.

\subsection{{\it Swift}} \label{sec:swift}
The X-Ray Telescope (XRT) on the Neil Gehrels Swift Observatory is a grazing-incidence focusing X-ray telescope, capable of detecting  photons in the energy range between 0.2 and 10 keV \citep{2004AIPC..727..637G, 2005_Burrows}. ToO observations of S3 1227+25 were performed with the \emph{Swift}-XRT between MJD~57158 and MJD~57180, which corresponds to midnight on May~16 until midnight on June~7, 2015 UTC in coordination with the VERITAS observations. A total observing time of 16.2 ks was accumulated during this period in photon-counting (PC) mode and the data were analyzed using the HEASOFT v6.21 package. 

The average photon count rate from the source in the 0.3--10 keV range for these observations was~$0.42~\pm~0.14$~cts/s. Photon pile-up at this rate level is negligible, and statistical uncertainties dominate over systematic ones. Photons were binned in energy and the resulting histogram was fit with an absorbed power-law spectrum assuming a Galactic column density of $N_{H}~= ~1.43 \times 10^{20}$ cm$^{-2}$. 
A good fit ($\chi^2 / \mathrm{ndf}$~=~39.4/40) is found for a power law normalization of $(2.73 \pm 0.09) \times 10^{-3}$ keV$^{-1}$ s$^{-1}$ at 1 keV and a photon index of $2.35 \pm 0.06$. A lightcurve of the absorption-corrected flux values is shown in Fig.~\ref{fig:mwl_lc} and the data are also used in the modeling of the broadband SED of the source in Section~\ref{sec:modeling}.

The Ultraviolet/Optical Telescope (UVOT) on the Neil Gehrels Swift Observatory is a photon-counting telescope. It is sensitive to photons with energies ranging roughly between 1.9 and 7.3 eV \citep{2005SS_Roming}.
\emph{Swift}-UVOT observations in the optical-UV band are performed in parallel to XRT measurements. \emph{Swift}-UVOT observations in the UW1, UW2, UM2, U, B, and V band filters were analyzed using the \texttt{uvotsource} routine by selecting source photons from a circular region of radius 5'' centered on S3 1227+25 and estimating the background in a 30''~radius circular region away from the blazar and containing no obvious sources in any band. The measured optical-UV fluxes are shown in Fig.~\ref{fig:mwl_lc} after correcting for interstellar extinction using the approach of \cite{2009_Roming}, assuming a reddening of  $E(B-V) = 0.0167$ \citep{2011_reddenning}.

\subsection{Optical} \label{sec:optical}

S3 1227+25 was observed with the 1.8~m Perkins telescope at the Lowell Observatory, and the 1.6~m Kuiper and 2.3~m Bok telescopes at the Steward Observatory in Arizona between May and July 2015. As seen in Fig.~\ref{fig:mwl_lc}, a low degree of polarization is observed in both data sets on the first night of observations. Observations and data reductions were performed in the same manner as described in \cite{2010_Jorstad} and \cite{2009_Smith}. The spectroscopic observations performed at Steward Observatory show no variation in the slope of the flux spectrum during the active period. Furthermore, S3 1227+25 is known to be highly variable at optical wavelengths, with the flux density ranging from 14 mag to 16 mag in the R band and the degree of polarization between 3\% and 25\% according to the optical monitoring at the Perkins telescope in Flagstaff, Arizona, USA. 

The All-Sky Automated Survey for Supernovae (ASAS-SN, \citealt{2014Shappee, 2017Kochanek}) is an automated program aimed at routinely surveying the entire visible sky for bright transient sources with minimal observational bias and reaches a depth of roughly 17 mag. ASAS-SN originally consisted of two stations, located at the Cerro Tololo International Observatory (CTIO, Chile) and the Haleakala Observatory (Hawaii) respectively. In 2017, three further stations were added, including a second unit at CTIO and one unit each at the McDonald Observatory (Texas) and the South African Astrophysical Observatory.

The observations made with ASAS-SN are available in two optical bands: the \emph{v} band,  centered at $\sim$ 551 nm and corresponding to observations from the two original stations, and the \emph{g} band, centered at $\sim$ 480 nm and corresponding to observations performed with the three new stations. 
For the purpose of this study, we obtain S3 1227+25 observations with ASAS-SN from the ASAS-SN Sky Patrol\footnote{\url{https://asas-sn.osu.edu/} (accessed on 02/27/2023)} for the time interval between MJD~57128 and MJD~57189, which correspond to midnight on April~16, 2015 until midnight on June~16, 2015 UTC, a month either side of the VERITAS detection.

The Asteroid Terrestrial-impact Last Alert System (ATLAS, \citep{ATLAS_main}) is a high cadence all sky survey system comprising two independent units, one on Haleakala (HKO), and one on Mauna Loa (MLO) in the Hawaiian islands. ATLAS was optimized to be an efficient system for finding potentially dangerous asteroids as well as in tracking and searching for variable and transient sources. It is capable of discovering more bright, less than 19 mag, supernovae candidates than other ground based surveys.
For the purpose of this study, we obtain S3 1227+25 observations with ATLAS in the R-band, centered at 679 nm, having a typical cadence of one data point per two days, shown in Fig.~\ref{fig:mwl_lc2}.


%

\subsection{Radio} \label{sec:radio}

Radio observations of S3 1227+25 at 15 GHz were routinely performed at a few-days cadence as part of the Owens Valley Radio Observatory (OVRO) \emph{Fermi}-LAT blazar monitoring program~\citep{2011ApJS..194...29R} and the lightcurve around the VERITAS detection is shown in Fig.~\ref{fig:mwl_lc}. The long-term lightcurve for the source, shown in Fig.~\ref{fig:mwl_lc2}, indicates that the blazar reached historically-high flux levels around the time of the VHE detection. 

Due to hardware issues, the monitoring was interrupted during the period surrounding the flare, with the last flux data point taken on MJD 57152 (five days before the \emph{Fermi}-LAT GeV flare detection and six days before the VERITAS one) and observations resumed on MJD 57229. The flux value recorded on MJD 57152 is used in the modeling of the broadband SED of the source in Section \ref{sec:modeling}, where the uncertainty on the value represents the range of fluxes measured 100 days around the date of the first VHE detection (MJD 57158).

Fig.~\ref{fig:mwl_lc} and Fig.~\ref{fig:mwl_lc2} show the 37 GHz observations of S3~1227+25 performed at the  Mets\"{a}hovi Radio Observatory with the 13.7~m radio telescope. This is located in Finland and at 37 GHz has a detection limit of the order of 0.2 Jy under optimal conditions. The typical integration time required to obtain a single flux density point is between 1200 and 1800 seconds and data points having a signal-to-noise ratio of less than 4 are considered as non-detections and shown as upper limits. The flux-density scale is set by observations of the HII region DR21 with the sources NGC 7027, 3C~274, and 3C 84 used as secondary calibrators. A detailed study of the methods used in the data analysis is found in \cite{1998_Teraesranta}. 

\section{Results}
\label{discussion}
\subsection{Temporal Variability}
\label{timescales}

Localizing the gamma-ray emission region in blazars is an indirect process with a variety of different methods previously used in the literature. The observation of rapid flux variability in these sources implies that the emission is coming from compact regions. Blazars have been found to show variability at timescales down to the order of a few minutes in both GeV and TeV bands, for example  PKS 2155-304 \citep{PKS_2155_minute_variability}, 3C 279 \citep{Ackermann_2016} and Mrk 421 \citep{2020_Mrk421}. It should be noted that compact emission regions do not automatically imply emission from near the SMBH as over-densities of the plasma can occur throughout the jet, including within the MT region. It has been proposed that these result from magnetic reconnection events (\citealt{RN18}, \citealt{giannios2}), the recollimation of the jet (\citealt{RN22}) or turbulence in the jet (\citealt{2021G_Marscher}).

The temporal gamma-ray variability of S3~1227+25 during the VERITAS detection was quantified by calculating the time taken for the flux to increase or decrease by a factor of 2. Known as the doubling or halving timescale, $\tau$, this is defined as:

\begin{equation}
    \centering
    F(t)=F(t_{0}) 2^{\tau^{-1}(t-t_{0})}
    \label{eq:4}
\end{equation}

where $F(t)$ and $F(t_{0})$ are the fluxes at times $t$ and $t_{0}$ respectively.
The measurement of doubling timescales is frequently used in the study of blazars  (for example \citealt{Foschini_2011}, \citealt{Saito_2013}) and applying the same method allows for comparability. A least squares fitting routine to Equation~\ref{eq:4} was applied to time intervals corresponding to flaring episodes in the 6 hour binned \textit{Fermi}-LAT lightcurve, shown in Fig. \ref{fig:mwl_lc}.

There is no general consensus on how to define a flaring episode (for example \citealt{RN50}, \citealt{2019_Meyer}). \cite{Nalewajko_2013} define flares as a contiguous interval of time containing a flux peak, having a flux higher than half the peak value of the entire observation. \cite{2019_Meyer} adopt a simple two-step procedure for identifying blocks of data points, having a flux higher than both the preceding and subsequent blocks and proceeding in both directions, as long as the blocks have successively lower fluxes. In this study, we define a flare using the method described in \cite{2021_Acharyya}, which combines these two approaches. We identify local peaks in flux, defined as bins having a higher flux than both the preceding and succeeding bin. We then proceed in both directions, as long as the corresponding bins have, within errors, successively lower or compatible fluxes. We also impose the condition that the peak of the flare must, within errors, have a flux greater than twice the average flux during the entire observation period (i.e $F_{\text{peak}} - \Delta F_{\text{peak}} \geq 2*F_{\text{mean}}$). Also shown in Fig. \ref{fig:mwl_lc}, are the two time periods satisfying our definition of a flare and used to compute the variability timescales.



The fastest observed variability timescale is found to be $\tau_{\text{obs}}~=~6.2~\pm~0.9$ hours. This corresponds to the time interval between $t_{0} = $ MJD~57178 and $t = $~MJD~57179.5, where the flux decreased from $F(t_{0})~=~(19.9~\pm~5.2)~\times~10^{-7}$ photons $\text{cm}^{-2}\text{s}^{-1}$ to  $F(t)~=~(1.8~\pm~0.9)~\times~10^{-7}$ photons $\text{cm}^{-2}\text{s}^{-1}$. Using geometric arguments, the shortest observed variability timescale was used to constrain the size of the emission region:

\begin{equation}
    \centering
    r \leq \frac{c \delta \tau_{\text{obs}}}{1+z}
	\label{eq:5}
\end{equation}
where $r$ is the size of the emission region, $c$ is the speed of light, the Doppler factor of the jet, $\delta~=~9.6$ \citep{Dopper_factor} and redshift of the source,  $z = 0.325$. The measured fastest observed variability timescale, $\tau_{\text{obs}}~=~6.2~\pm~0.9$~hours, corresponds to a gamma-ray emission region of size $r~\leq~(6.4~\pm~0.9)~\times~10^{13}$~m. It should be noted that the Doppler factor of the jet, $\delta~=~9.6$, measured in \cite{Dopper_factor} is obtained from archival SEDs, which may not be contemporaneous datasets and a variability timescale in the host frame of 1 day is considered for all sources, with the Doppler factor assumed to be equal to the Lorentz factor.

\subsection{Gamma-ray spectrum}
\label{photon-photon-pair-production}

\begin{table*}
	\centering
	\caption{The best fit spectral parameters obtained from a joint fit to the combined gamma-ray spectrum for S3 1227+25, observed with the \textit{Fermi}-LAT and VERITAS between May 16 and May 18, 2015 UTC (MJD 57158 -- 57160), using PL, PLE and LP functions. The final two columns state the  $\chi^{2}/\text{n.d.f}$ and AIC values obtained for each model.}
    \hspace{-2 cm}
	\label{tab:table4}
	\resizebox{ \textwidth}{!}{
	\centering
	\begin{tabular}{lccccccr} 
            \hline
		Fit  &$N_{0}$ &$\alpha$ &$\beta$ &$E_0$  &$E_{\text{cut}}$ & $\chi^{2}/\text{n.d.f}$ &$\text{AIC}$ \\
		function &[$ 10^{-12} \text{cm}^{-2}\text{s}^{-1} \text{MeV}^{-1}$] & & &[MeV] & [GeV] & &\\
		\hline
		PL &24.8 $\pm$ 4.3 &  2.23 $\pm$ 0.04 &- &816 &-  &1.50 &6.27\\
		PLE &28.0 $\pm$ 5.2 & 2.10 $\pm$ 0.10 &-  &816 &282 $\pm$ 203 &1.40 &6.22\\
		LP &30.2 $\pm$ 7.4 & 2.07 $\pm$ 0.16 &0.04 $\pm$ 0.03   &816 &- &1.53 &7.19\\
		\hline
	\end{tabular}}
\end{table*}

Regions near the central engine of AGN are photon-rich environments and the interactions between these photons and gamma-ray photons results in photon-photon pair production ($\gamma \gamma \rightarrow e^{+} e^{-}$). The MT has comparatively a much lower photon density than the BLR, meaning there is less likelihood of pair production within the MT. Pair production manifests itself as an attenuation of the gamma-ray spectrum for emission coming from near the SMBH \citep{RN2,Stern_2014}, whereas emission originating from within the MT is not expected to have this spectral feature (\citealt{Donea_2003}, \citealt{RN28}). It should be noted, that the presence of a potential cut-off in the spectrum can also be the consequence of a break in the energy distribution of the emitting electrons (\citealt{Dermer_2015}). 

To investigate the spectral attenuation, we corrected the VERITAS and \textit{Fermi}-LAT flux points for attenuation by extragalactic background light (EBL) using the \cite{RN10} model and assuming redshift  $z = 0.325$. A joint fit was performed to the de-absorbed gamma-ray spectrum for the time interval corresponding to the VERITAS detection and is shown in Fig. \ref{fig:combined_spectrum}. The combined gamma-ray spectrum was modeled with a simple power law (PL) model of the form $\frac{dN}{dE}=N_{0} \left(\frac{E}{E_0}\right)^{-\alpha}$, a power law function with exponential cut-off (PLE) model of the form $\frac{dN}{dE}=N_{0} \left(\frac{E}{E_0}\right)^{-\alpha}  e^{-\left(\frac{E}{E_{\text{cut}}}\right)}$ and a log parabolic (LP) model of the form $\frac{dN}{dE}=N_{0} \left(\frac{E}{E_0}\right)^{-\alpha - \beta \text{ln} \left(\frac{E}{E_0}\right)}$. Here  $N_{0}$ is the normalization, $\alpha$ is the spectral index, $\beta$ the spectral curvature, $E_0$ the pivot energy and $E_{\text{cut}}$ the cut-off energy. The best fit spectral parameters are tabulated in Table \ref{tab:table4}, along with the $\chi^{2}/\text{n.d.f}$ obtained for each fit. It should be noted that despite covering the same time period, the \textit{Fermi}-LAT and VERITAS spectral data are not strictly contemporaneous.

To compare the fits provided by all three models an Akaike Information Criterion (AIC) test  \citep{RN36} was performed. The AIC of a model $s$ is given by, $\text{AIC}_{s} = - 2 \text{ln}  {L}_{{s}} + 2 k_{f_{s}}$, where ${L}_{{s}}$ is the likelihood of the model $s$ and ${k}_{{f}_{{s}}}$ is the number of free parameters in the model. The difference in AIC values between two models s and s', $\Delta \text{AIC}_{\text{s},\text{s'}}= \text{AIC}_{\text{s}}- \text{AIC}_{\text{s'}}$, estimates how much more the model s diverges from the true distribution than the model s' (\citealt{RN52}, \citealt{RN51}). A lower AIC value means a better description of the spectrum and an AIC difference of greater than 2 between two models means that the model with the higher AIC is significantly worse than the model with the lower AIC value (\citealt{RN21}). The AIC values for the PL, PLE and LP models are tabulated in Table \ref{tab:table4}. The combined gamma-ray spectrum was found to favor neither of the three models significantly, with $\Delta \text{AIC}_{\text{PL},\text{PLE}} = 0.05$ and $\Delta \text{AIC}_{\text{LP},\text{PLE}} = 0.97$ respectively. 



\subsection{Results from multiwavelength correlation analysis}

In SSC models of blazar emission, both synchrotron and IC scattering derive from the same population of electrons, resulting in strong correlations between low and high energy wavebands \citep{1994_Sikora}. On the other hand, the observation of orphan flares in one waveband having no correlation with others, for example the 2002 flare of 1ES 1959+650 \citep{2004_orphan}, can be interpreted as evidence of multiple emission zones or support for hadronic emission models. Investigating multiwavelength correlations helps in identifying relationships between emission zones and in placing constraints on the dominant mechanisms responsible.

\begin{figure*}[t!]
\centering
\includegraphics[width= \linewidth]{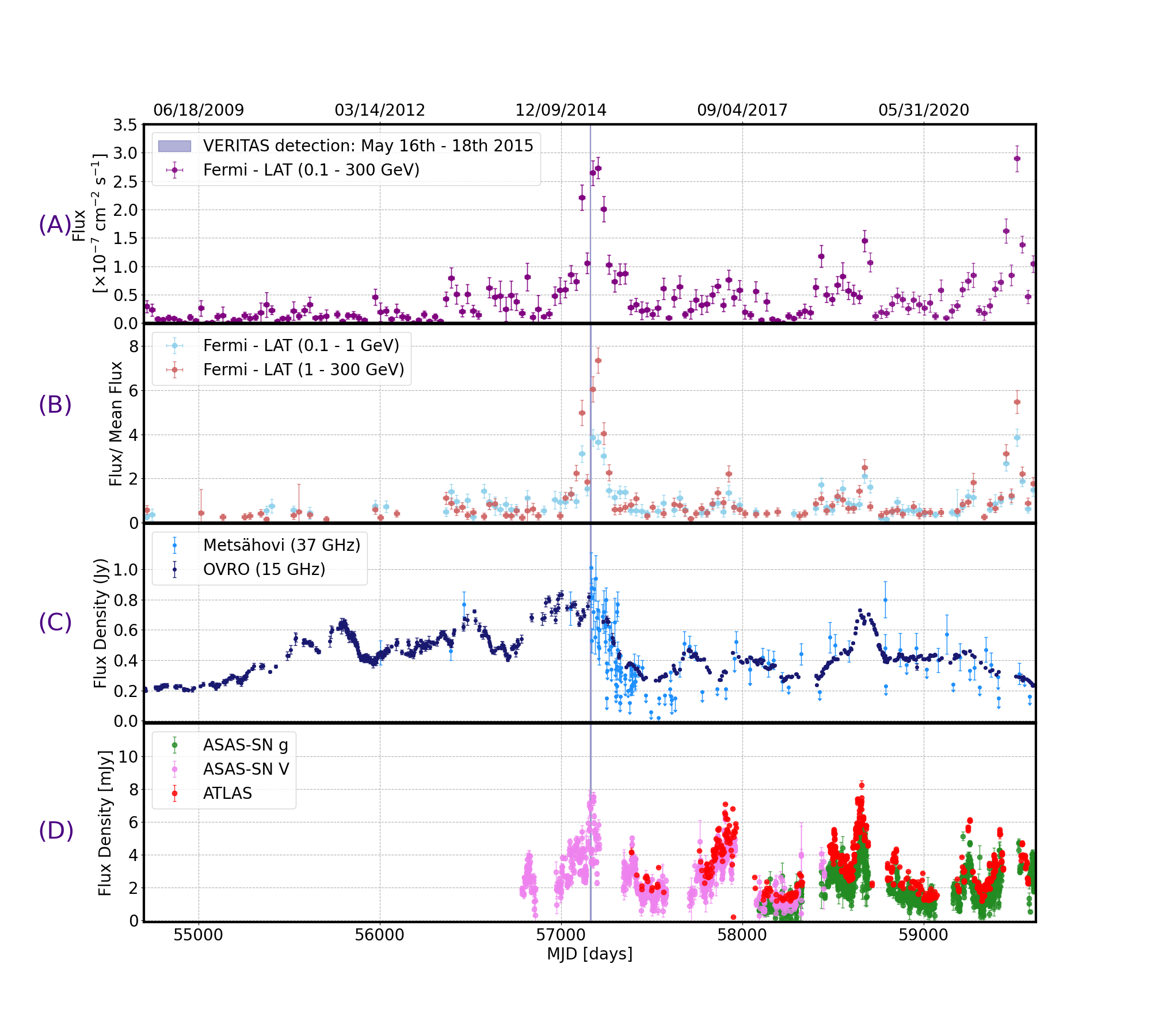}
\caption{Multiwavelength lightcurves of S3 1227+25 over a wider time interval, MJD~54683 to MJD~59611, corresponding to midnight on August 4, 2008 until midnight on February 1, 2022. The blue narrow band corresponds to the VERITAS detection. \textbf{(A)} \emph{Fermi}-LAT lightcurve in the 0.1--300 GeV band.  \textbf{(B)} Separate \emph{Fermi}-LAT lightcurves showing the low energy flux (0.1 $\leq$ $\text{E}_{\gamma}$ $\leq$ 1 GeV) as blue circles and the high energy flux (1 $\leq$ $\text{E}_{\gamma}$ $\leq$ 300 GeV) using red circles. To aid visual comparison, the individual flux values have been divided by the mean flux in the corresponding energy ranges. The error bars are purely statistical and only data points with TS $\geq$ 10 are shown. \textbf{(C)} Radio flux observations: OVRO (dark blue) and Mets\"{a}hovi (light blue) lightcurves at 15 GHz and 37 GHz respectively. For the Mets\"{a}hovi data, individual bins having a signal-to-noise ratio of less than 4 are considered non-detections and shown as upper limits. \textbf{(D)} Optical flux observations: Optical data, not corrected for extinction, from the ASAS-SN Sky Patrol in the g and V band  (green and magenta respectively) and ATLAS R band data (in red).}
\label{fig:mwl_lc2}
\end{figure*}

In Fig.~\ref{fig:mwl_lc}, we present multiwavelength lightcurves corresponding to the time period around the VERITAS detection. We now consider multiwavelength lightcurves over a wider time interval, MJD~54683 to MJD~59611 and discuss potential correlations between them. This corresponds to midnight on August 4, 2008, the start of the \textit{Fermi}-LAT mission until midnight on February 1, 2022. The lightcurves considered are presented in Fig. \ref{fig:mwl_lc2} and include the \textit{Fermi}-LAT lightcurve, obtained using the data reduction routine described in Section \ref{sec:fermi}, as well as radio lightcurves, measured with Mets\"{a}hovi and OVRO and optical lightcurves, obtained from the ASAS-SN Sky Patrol and ATLAS data. There were not sufficient statistics to enable a cross-correlation analysis using the VERITAS and X-ray observations. The \textit{Fermi}-LAT data are also investigated in two distinct energy ranges: 0.1--1 GeV (low energy) and 1--300 GeV (high energy), binned in daily intervals using the procedure outlined in Section \ref{sec:fermi} and shown in Fig~\ref{fig:mwl_lc2}.

Local cross-correlation functions (LCCFs; \citealt{Welsh_LCCF}) are applied to pairs of two different lightcurves to search for correlations between them. For two lightcurves having fluxes $a_{i}$ and $b_{j}$, corresponding to times $t_{a_{i}}$ and $t_{b_{j}}$, the LCCF is computed as:

\begin{equation}
    \centering
    \text{LCCF} (\tau) = \frac{1}{M}\frac{\Sigma (a_{i} - \overline{a}_{\tau}) (b_{j} - \overline{b}_{\tau})}{\sigma_{a \tau}\sigma_{b \tau}}
    \label{eq:LCCF}
\end{equation}

where the sums run over M pairs for which $\tau \leq t_{a_{i}}-t_{b_{j}}<\tau+\Delta~t$ for a chosen timestep $\Delta~t$, $\overline{a}_{\tau}$ and $\overline{b}_{\tau}$ are flux averages and $\sigma_{a \tau}$ and $\sigma_{b \tau}$ are standard deviations over the M pairs respectively \citep{Welsh_LCCF}. As also adopted in \cite{2019_Meyer}, for the binning of the timelags, $\tau$, we choose the higher half median of the time separations between consecutive data points in the two lightcurves. Furthermore, the minimum and maximum values of $\tau$ are chosen to be $\pm 0.5$ times the length of the shorter lightcurve \citep{Max_Moerbeck_CCFs}.

\begin{figure*}
    \centering
    \resizebox{\linewidth}{!}{
    \includegraphics{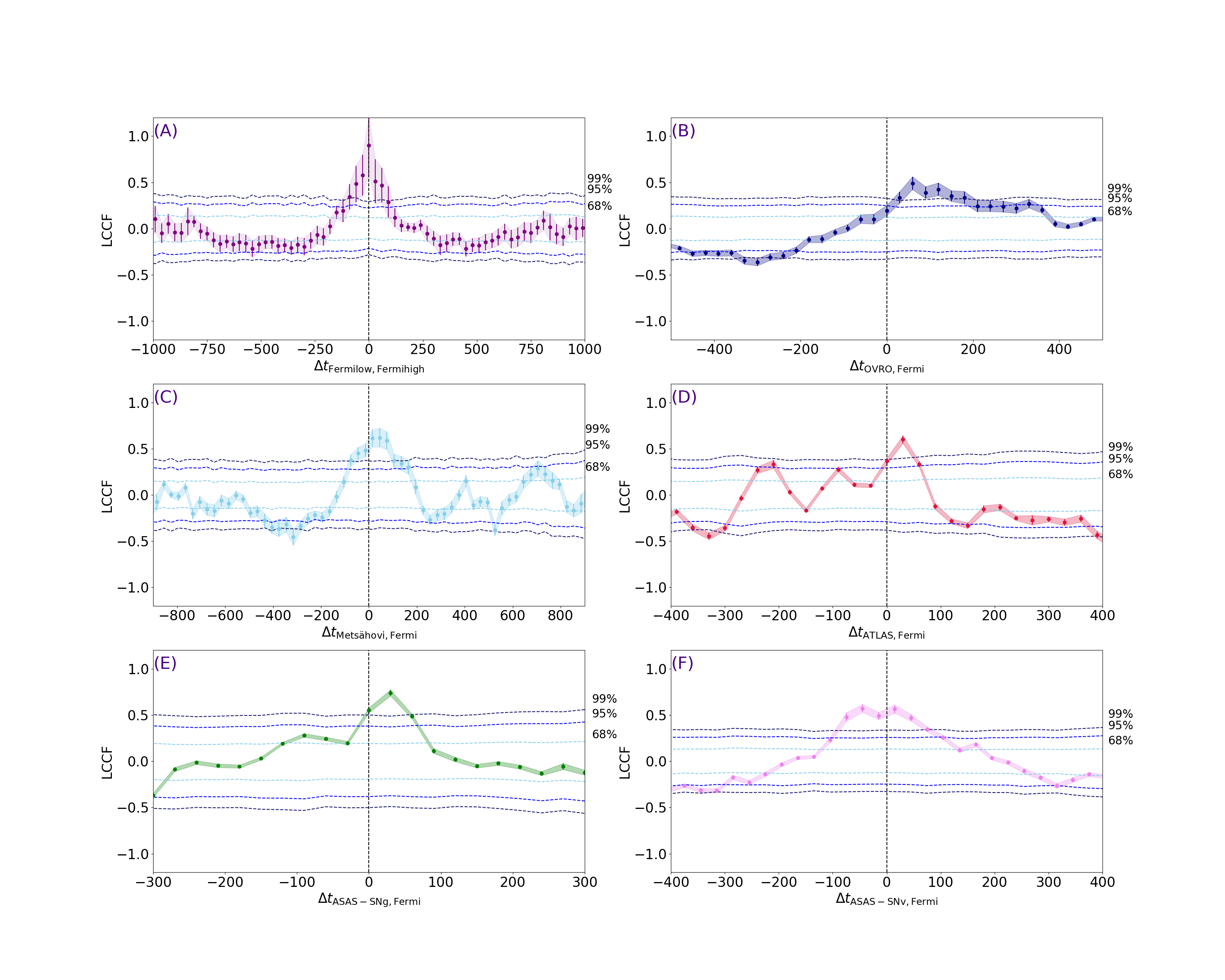}}
    \caption{Local cross-correlation functions (LCCFs) calculated between the different lightcurves are shown in color. The x-axis represents the time delay, $\Delta t _{a,b} = t_{a} -t_{b}$, in days for two lightcurves $a$ and $b$. The dashed  vertical line corresponds to $\Delta t _{a,b} = 0$. The corresponding shaded regions indicate the error bounds of the LCCFs. The blue lines represent the 68$\%$, 95$\%$ and 99$\%$ confidence intervals (from lighter to darker shades) derived from Monte Carlo simulations.}
    \label{fig: Fig 5.}
\end{figure*}

The LCCF method is independent of any differences in sampling rates of the two lightcurves. Furthermore, unlike Discrete Correlation Functions (DCFs; \citealt{RN19}), LCCFs are intrinsically bound in the interval [-1,1]. Moreover, the use of LCCFs has been found to be more efficient than DCFs in the study of correlations (\citealt{Max_Moerbeck_CCFs}). The LCCF obtained between the low and high energy \textit{Fermi}-LAT lightcurves is shown in Fig. \ref{fig: Fig 5.}, along with the LCCFs obtained between the full energy range \textit{Fermi}-LAT lightcurve and each of the other lightcurves considered. 

Monte Carlo simulations are used to determine the significance of an obtained LCCF value, using 1000 artificial light curves that match the probability  distribution function (PDF) and power spectral density (PSD) of each observation using the method outlined in \cite{Emmanoulopoulos}. The 68$\%$, 95$\%$ and 99$\%$ confidence intervals obtained are shown in Fig.~\ref{fig: Fig 5.}. As seen in panel (A) of Fig. \ref{fig: Fig 5.}, the \textit{Fermi}-LAT lightcurves in the different energy regimes are found to show evidence of a strong correlation with a peak compatible with zero time lag. This can be interpreted as evidence that the MeV and GeV components of the flare have the same origin. 

Panels (B) and (C) of Fig.~\ref{fig: Fig 5.} show the correlations between the \textit{Fermi}-LAT data in the 0.1--300 GeV energy band and the 37 GHz and 15 GHz radio flux densities, as measured with Mets\"{a}hovi and OVRO respectively. In both cases, a positive correlation is found, significant at a level of above 3$\sigma$, and consistent with zero time lag. Moreover, as seen in Fig.~\ref{fig:mwl_lc2}, the radio outburst also started several years earlier than the gamma-ray flare. This could imply that the multi-wavelength outburst was built in the jet and the gamma-ray flare and VHE detection were the culmination of it. Furthermore, while  we observe visual evidence of a delay between the 37 GHz and 15 GHz radio observations with respect to the \textit{Fermi}-LAT lightcurve, it should be noted that the OVRO observations were interrupted during the period surrounding the flare, with the last flux data point taken on MJD 57152 (five days before the \emph{Fermi}-LAT GeV flare detection) due to hardware issues and resumed on MJD 57229. There was not enough data to investigate if the delay,  as seen in the low-state data, is also found during the VHE detection period and it is not considered for further discussion.

Panels (D), (E) and (F) of Fig.  \ref{fig: Fig 5.} show the correlations between the \textit{Fermi}-LAT data in the 0.1--300 GeV energy range and the optical data taken with ATLAS in the R band and ASAS-SN observations in the g and V bands respectively. A positive correlation, significant at a level of above 3$\sigma$ and consistent with zero time lag is seen for correlations between \textit{Fermi}-LAT data and ATLAS and ASAS-SN g data, while we find evidence of a broader peaked correlation, also compatible with a zero lag, for correlations between \textit{Fermi}-LAT and ASAS-SN V observations. The strong optical and gamma-ray correlations found for S3 1227+25 suggest a single-zone model of emission. Under the assumption that the optical and gamma-ray flares are produced by the same outburst propagating down the jet, both positive and negative time-lags on timescales of days are predicted in SSC and EIC models (for example \citealt{Sokolov_2004}).

\subsection{SED Modeling}
\label{sec:modeling}

The final method that sheds light on the mechanisms responsible for the VERITAS detection of S3 1227+25 is  modeling the multiwavelength SED. In a leptonic scenario, the second peak in the broadband SED is interpreted as SSC emission or EIC emission (for example \citealt{1992_Maraschi, 1994_Sikora}). We model the multiwavelength broadband SED of S3 1227+25 taken between May 16 and May 18, 2015 using \emph{agnpy}, an open-source Python package for modeling radiative processes involving relativistic particles accelerated in the jets of AGN \citep{2022_Nigro}. The synchrotron and SSC processes in \emph{agnpy} are based on the work published in \cite{2008_Finke} and \cite{2009_dermer_menon}. We assume a 20$\%$ systematic uncertainty on the VHE flux measurements, 10$\%$ on the \emph{Fermi}-LAT and X-ray band data and 5$\%$ on the lower energy UV band. The radio SED point from OVRO was not considered in the fit due to a larger integration region, contaminated by extended jet emission.


\begin{figure*}[t!]
\centering
\includegraphics[width=0.6 \linewidth]{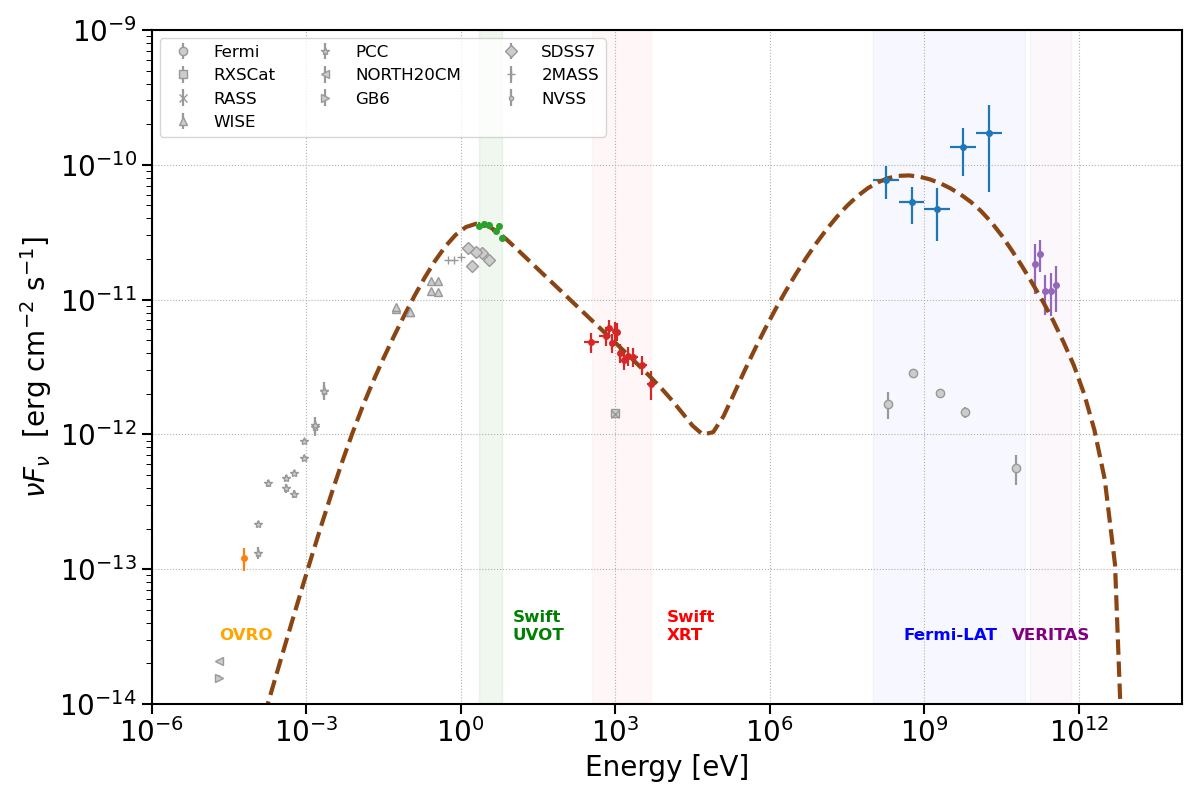}
\caption{The multiwavelength broadband SED of S3 1227+25 during the nights of the VERITAS detection, May 16 -- May 18, 2015 (MJD 57158 -- 57160), fitted using \emph{agnpy}. The grey points represent archival data. The dashed brown curve represents the best fit one-zone synchrotron and SSC radiation model, obtained using all of the 5000 samples from all of the 30 walkers, with the parameters shown in Table~\ref{tab:table1}.}
\label{fig:sed}
\end{figure*}

We use a simple one-zone synchrotron and SSC radiation model consisting of a single blob with a broken power law electron energy density (EED). This scenario is most commonly considered to model BL Lacs like S3 1227+25. It consists of nine parameters, including three describing the blob, namely the Doppler factor, $\delta$, the magnetic field $B$, and the radius, $r$, constrained from the shortest observed variability timescale for this time period (see Section \ref{timescales}). The other six parameters define the EED and are the electron density, $k_{e}$, the initial and final spectral indices, $p1$ and $p2$, the break Lorentz factor of the electron distribution, $\gamma_{\text{b}}$, and the minimum and maximum Lorentz factors of the EED, $\gamma_{\text{min}}$ and $\gamma_{\text{max}}$ respectively. More details of the parameters can be found in \cite{2022_Nigro}. 

The parameters $\delta$, $r$, $\gamma_{\text{min}}$ and $\gamma_{\text{max}}$ are kept fixed while the other parameters are left free during the fitting procedure. To limit the span of the parameter space, we fit the $\text{log}_{10}$ of the parameters whose range is expected to cover several orders of magnitude, namely $k_{e}$, $\gamma_{\text{b}}$ and $B$. The fit is performed using the Markov Chain Monte Carlo (MCMC) method \citep{gamerman_mcmc,hogg_mcmc}. The MCMC sampling was run for 5000 steps with 30 walkers (six times the number of free parameters), with the first 2000 steps discarded as a burn-in period. The values of the parameters are reported in Table~\ref{tab:table1}.  As seen in Fig.~\ref{fig:sed}, a simple one-zone synchrotron and SSC radiation model describes the data well with a good agreement seen between the fitted model and the data over the energy range considered. 

\begin{table}
\centering
\caption{The best fit parameters of a simple one-zone synchrotron and SSC radiation model fitted to the multiwavength broadband SED of S3 1227+25 taken between May 16 and May 18, 2015 (MJD 57158 -- 57160). The parameters $\delta$, $r$, $\gamma_{\text{min}}$, $\gamma_{\text{max}}$ and redshift, $z$, have been kept fixed while the other parameters are left free during the fitting procedure. The best-fit model is shown in Figure \ref{fig:sed}.}
\label{tab:table1}
\begin{tabular}{lr} 
\hline
parameter &value \cr
\hline

 $\text{log}_{10} ({\text{B [G]}})$ &$0.10^{+0.03}_{-0.02}$ \cr
 $\text{log}_{10} (k_{e} [\text{cm}^{-3}])$ &$-2.23^{+0.11}_{-0.12}$\cr
 $\text{log}_{10} (\gamma_{\text{b}})$ &$3.51^{+0.04}_{-0.04}$ \cr
 $p1$ &$1.40^{+0.41}_{-0.28}$  \cr
 $p2$ &$3.74^{+0.02}_{-0.02}$ \cr
 $\delta$ &9.6 \cr 
 $r [10^{13} m]$ &6.40 \cr
 $z$ &0.325 \cr
 $\gamma_{\text{min}}$ &500 \cr
 $\gamma_{\text{max}}$ &$10^{6}$\cr
 \hline
\end{tabular}
\end{table}

\section{Conclusions} 
\label{sec:conclusions}

In this work, we report the follow-up to the discovery of VHE emission from S3 1227+25 with VERITAS and  undertake a temporal and spectral analysis of the multiwavelength data during this period, in particular the gamma-ray observations. The shortest observed variability timescale obtained with the \textit{Fermi}-LAT for S3~1227+25 during the VERITAS detection is $\tau_{\text{obs}}=6.2 \pm 0.9$ hours. 
A joint fit to the combined gamma-ray spectrum was found to favor neither a power law, log parabolic or power law with an exponential cut-off model significantly. 

The cross-correlation studies between the gamma-ray and radio lightcurves show a positive correlation, significant at a level of above 3$\sigma$, and consistent with zero time lag. Moreover, a positive correlation, significant at a level of above 3$\sigma$ and consistent with zero time lag is also seen between \textit{Fermi}-LAT data and ATLAS and ASAS-SN g data, while we find evidence of a broader peaked correlation, also compatible with a zero lag, between \textit{Fermi}-LAT and ASAS-SN V observations. The final method that sheds light on the mechanisms responsible for the VERITAS detection of S3~1227+25 was modeling the multiwavelength SED taken between May~16 and May~18,~2015. We consider a simple one-zone synchrotron and SSC radiation model consisting of a single blob accelerating with a broken power law EED. A good agreement was seen between the fitted model and the multiwavelength spectral data over the entire energy range considered.

The results of all the investigations, put together, lead to the natural conclusion that a one-zone SSC model explains the gamma-ray observations of S3 1227+25 during the VERITAS detection. There is evidence to suggest a compact emission region, namely the short variability timescales. Furthermore, the cross-correlation studies between the different lightcurves, in particular the strong optical and gamma-ray correlations, also suggest a single-zone model of emission. This result is further reinforced by evidence that the multiwavelength SED is well described by a simple one-zone leptonic SSC radiation model.

\begin{acknowledgments}
\section*{Acknowledgments}

This research is supported by grants from the U.S. Department of Energy Office of Science, the U.S. National Science Foundation and the Smithsonian Institution, by NSERC in Canada, and by the Helmholtz Association in Germany. This research used resources provided by the Open Science Grid, which is supported by the National Science Foundation and the U.S. Department of Energy's Office of Science, and resources of the National Energy Research Scientific Computing Center (NERSC), a U.S. Department of Energy Office of Science User Facility operated under Contract No. DE-AC02-05CH11231. We acknowledge the excellent work of the technical support staff at the Fred Lawrence Whipple Observatory and at the collaborating institutions in the construction and operation of the instrument. 

This work has made use of data from the Asteroid Terrestrial-impact Last Alert
System (ATLAS) project. The Asteroid Terrestrial-impact Last Alert System
(ATLAS) project is primarily funded to search for near earth asteroids through
NASA grants NN12AR55G, 80NSSC18K0284, and 80NSSC18K1575; byproducts of
the NEO search include images and catalogs from the survey area. This work was
partially funded by Kepler/K2 grant J1944/80NSSC19K0112 and HST GO-15889,
and STFC grants ST/T000198/1 and ST/S006109/1. The ATLAS science products
have been made possible through the contributions of the University of Hawaii
Institute for Astronomy, the Queen’s University Belfast, the Space Telescope
Science Institute, the South African Astronomical Observatory, and The
Millennium Institute of Astrophysics (MAS), Chile.

This study was partly based on observations conducted using the 1.8 m Perkins Telescope Observatory (PTO) in Arizona (USA), which is owned and operated by Boston University.
The BU group was supported in part by NASA Fermi Guest Investigato grant 80NSSC22K1571.

This publication has made use of data obtained at Metsähovi Radio Observatory, operated by Aalto University in Finland.

This research has made use of data from the OVRO 40-m monitoring program \citep{2011ApJS..194...29R}, supported by private funding from the California Institute of Technology and the Max Planck Institute for Radio Astronomy, and by NASA grants NNX08AW31G, NNX11A043G, and NNX14AQ89G and NSF grants AST-0808050 and AST-1109911.

This work has made use of data from the Steward Observatory, supported by NASA Fermi Guest Investigator grant NNX12AO93G.

Part of this work is based on archival data, software or online services provided by the Space Science Data Center - ASI.

A.A. and M.S. acknowledge support through NASA grants 80NSSC22K1515, 80NSSC22K0950, 80NSSC20K1587, 80NSSC20K1494 and NSF grant PHY-1914579. 

S.K. acknowledges support from the European Research Council (ERC) under the European Unions Horizon 2020 research and innovation programme under grant agreement No.~771282.

\end{acknowledgments}

\bibliography{s3_1227}{}
\bibliographystyle{aasjournal}



\end{document}